# Self-assembled chiral phosphorus nanotubes from phosphorene : a molecular dynamics study

Douxing Pan[1,2,3], Tzu-Chiang Wang[2,*], Chao Wang[2], Wei Guo[1], and Yugui Yao[1,*]

[1]Laboratory of quantum functional material design and application, School of Physics, Beijing Institute of Technology, Beijing 100081, China

[2]State Key Laboratory of nonlinear Mechanics, Institute of Mechanics, Chinese Academy of Sciences, Beijing 100190, China

[3]University of Chinese Academy of Sciences, Beijing 100190, China

[*]Address correspondence to Tzu-Chiang Wang: tcwang@imech.ac.cn; ygyao@bit.edu.cn (YGY)

Controlled syntheses in nanoscale structures should be expected and phosphorous nanotubes with predefined chiralities are important in electronic devices with tunable bandgap. Herein, incorporating molecular dynamics simulations with theoretical analyses, we show that a zigzag phosphorene nanoribbon can self-assemble and form a corresponding chiral phosphorous nanotube surrounding a template armchair phosphorous nanotube. The van der Waals potential between the nanoribbon and the nanotube is transformed to the intrinsic deformed and chemical bonding energies of the synthesized tube together with partial kinetic energy. The self-assembly process has an apparent temperature dependence and size effect, also, the formed chiral tube is thermodynamically stable and its chirality and measurement can be tuned by the radius of the template tube and the aspect ratio of the raw ribbon. The study suggests a novel and feasible approach for controlled synthesis of phosphorous nanotubes and thus is of great interest for semiconductor device applications.

The mechanical behaviors of two-dimensional (2D) materials are crucial to a variety of applications ranging from ultrafast electronics to multifunctional nanocomposites[1,2]. Phosphorene, as a fire-new graphene-like single-element 2D material, is promising for thin-film electronics, infrared optoelectronics and novel devices where anisotropic properties are desirable[3,4,5]. The corrugated crystal structure of phosphorene not only leads to novel physical properties, but also produces anisotropic mechanical behaviors[4,5,6], for example, apparent nonlinear Young's modulus and ultimate strain[6], prominent anisotropic elasticity[7], negative Poisson's ratio[8] and highly anisotropic ripple pattern[4] and bending-induced extension[5]. Recently, the fracture behaviors of phosphorene at finite temperatures are explored and the failure crack propagates preferably along the groove in the corrugated structure when uniaxial tension is applied in its armchair direction[9]. Moreover, as typical phosphorene allotropes, the structural stability and mechanical properties of phosphorus nanotubes (PNTs), and Young's modulus and fracture strength are found to be strongly dependent on the diameter and length of the nanotube[10].

Due to the tremendous needs for functional nanomaterials[2,11,12], controlled synthesis should be crucial in nanoscale structures, and PNTs with predefined chiralities are desired for electronic devices with tunable bandgap. On the other hand, due to the realization of graphene with tailored properties using different techniques[13,14,15,16], self-assemblies of graphene ribbons (GNRs) have been extensively investigated, theoretically and experimentally[2,11,12,13]. Also, with the ever-increasing controllabilities in supramolecular organizations[2,12,14,15], GNRs have been proposed for the novel generation of nanostructures including carbon nanoscrolls (CNSs)[2,17,18], twisted GNRs and chiral CNTs[12,15], and versatile configurations assembled by fullerenes and graphene flakes including disordered, parallel and vertical wrappings[19]. However, Can the self-assembly happen for double-atomically thin phosphorene? If yes, can chiral phosphorus nanotubes (cPNTs)[20] be assembled from phosphorene with tailored shapes and sizes, and how?

In this work, we demonstrate by classic molecular dynamics (MD) simulations the self-assembly mechanical behaviors of phosphoene nanoribbons (PNRs) on the surface of PNTs. The study shows that an armchair phosphorous nanotube (aPNT) can induce a zigzag phosphorene nanoribbon (zPNR) to self-assemble into a corresponding cPNT. The van der Waals (vdW) interaction between the aPNT and zPNR is the driving force of self-assembly behavior. The self-assembly is not only dependent on the radius of template tube and size of raw ribbon, but also on the temperature. The formed chiral tube is thermodynamically stable and its chirality can be well controlled theoretically. Our studies provide a novel and feasible approach for the fine controlled synthesis of phosphorene allotropes and novel phosphorene-based composite functional nanostructures, and should be of great importance to developments of electronic devices with tunable bandgap.

The computational model is shown in Fig. 1, consisting of a zigzag phosphorene nanoribbon with a length $L_x$, width $L_y$ and thickness $h$ and a template armchair phosphorus nanotube with an equivalent radius $r$. The aPNT with a length at least $2L_y$, is initially fixed at a distance of $1.5r$ away from left side of the zPNR and around 0.35nm above the zPNR along z-direction which is the equilibrium distance between the aPNT and zPNR. Three different aPNTs with radii $r$=1.10,1.47 and 1.84nm, are used to explore the curvature of the template aPNT on the self-assembly behavior of the zPNR. Similar to the slenderness ratio of the column for characterization of size effect in mechanics of materials[21], an aspect ratio of the zPNR is introduced as

$$\delta = \frac{L_x}{L_y} = \frac{m|\boldsymbol{a}_2|}{n|\boldsymbol{a}_1|} = \alpha \frac{m}{n} \tag{1}$$

where lattice vectors $\boldsymbol{a}_1$ and $\boldsymbol{a}_2$ are along armchair $(n,0)$ and zigzag $(0,m)$ directions of phosphorene, respectively, and $n$, $m$ are number of unit cells along the two typical directions, as denoted in Fig.1. $\alpha=|\boldsymbol{a}_2|/|\boldsymbol{a}_1|$ is a dimensionless parameter and can be viewed as $\alpha=1$ for convenience.

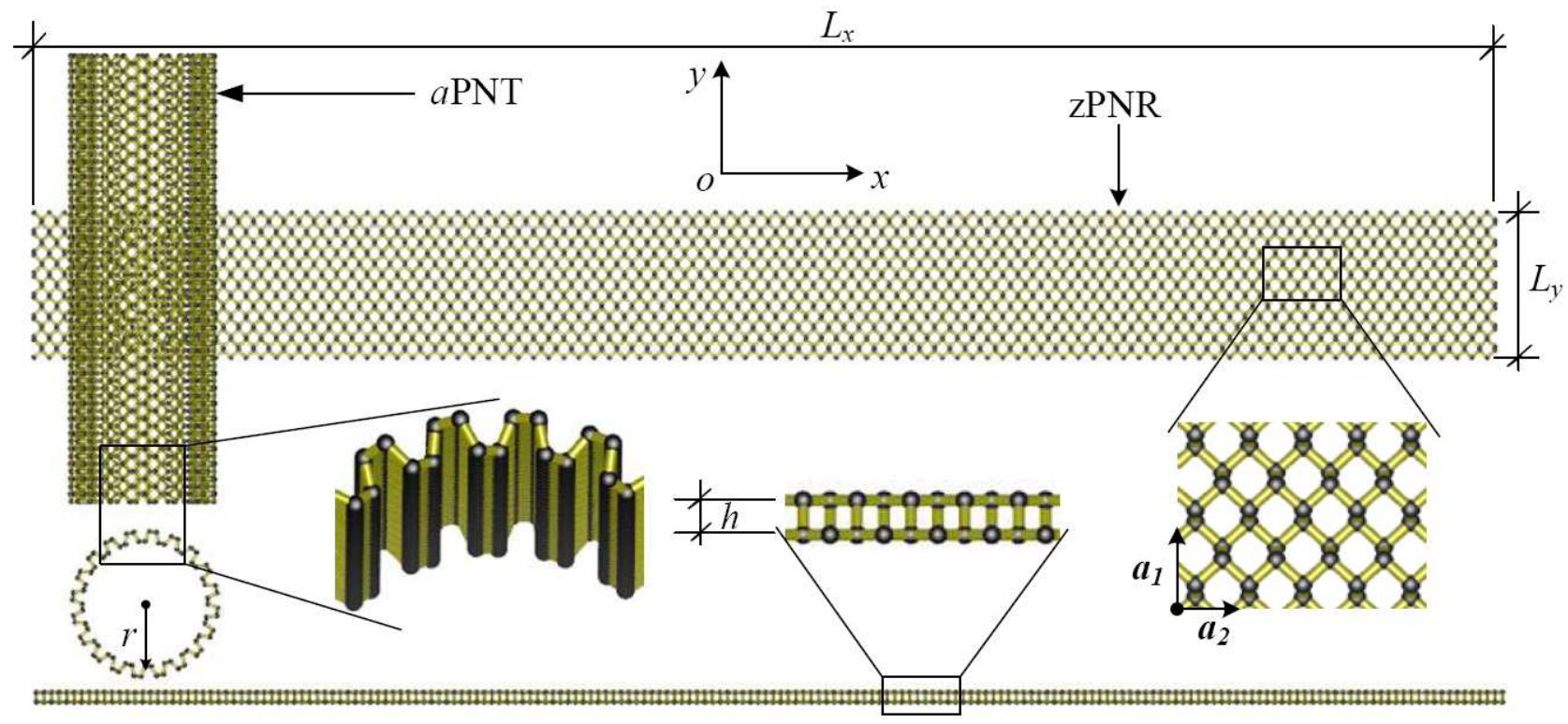

**Fig. 1.** (Color online) Computational model of zPNR and aPNT. $L_x$, $L_y$ and $h$ are length, width and thickness of zPNR, respectively, and $r$ is an average radius of aPNT. Lattice vectors $\boldsymbol{a}_1$ and $\boldsymbol{a}_2$ are along armchair$(n,0)$ and zigzag$(0,m)$ directions, respectively.

All simulations were carried out within the framework of classic MD-based method, implemented in the large-scale atomic/molecular massively parallel simulator (LAMMPS) code[22]. The interatomic interactions in aPNT or zPNR are characterized by the Stillinger–Weber (SW) potential[23], which has been previously parameterized to correctly describe the mechanical and thermal properties of phosphorene by Jiang *et. al.* [24] and Xu *et. al.* [25], and cited by the literatures mentioned in the introduction[6,9,10]. The interaction between aPNT and zPNR is described by the Lennard-Jones (LJ)

potential

$$V_{LJ}(r) = 4\varepsilon\left[\left(\sigma/r\right)^{12} - \left(\sigma/r\right)^{6}\right] \tag{2}$$

where $\varepsilon$ = 0.01594 eV and $\sigma$ = 0.3438 nm for bilayer phosphorene[26].

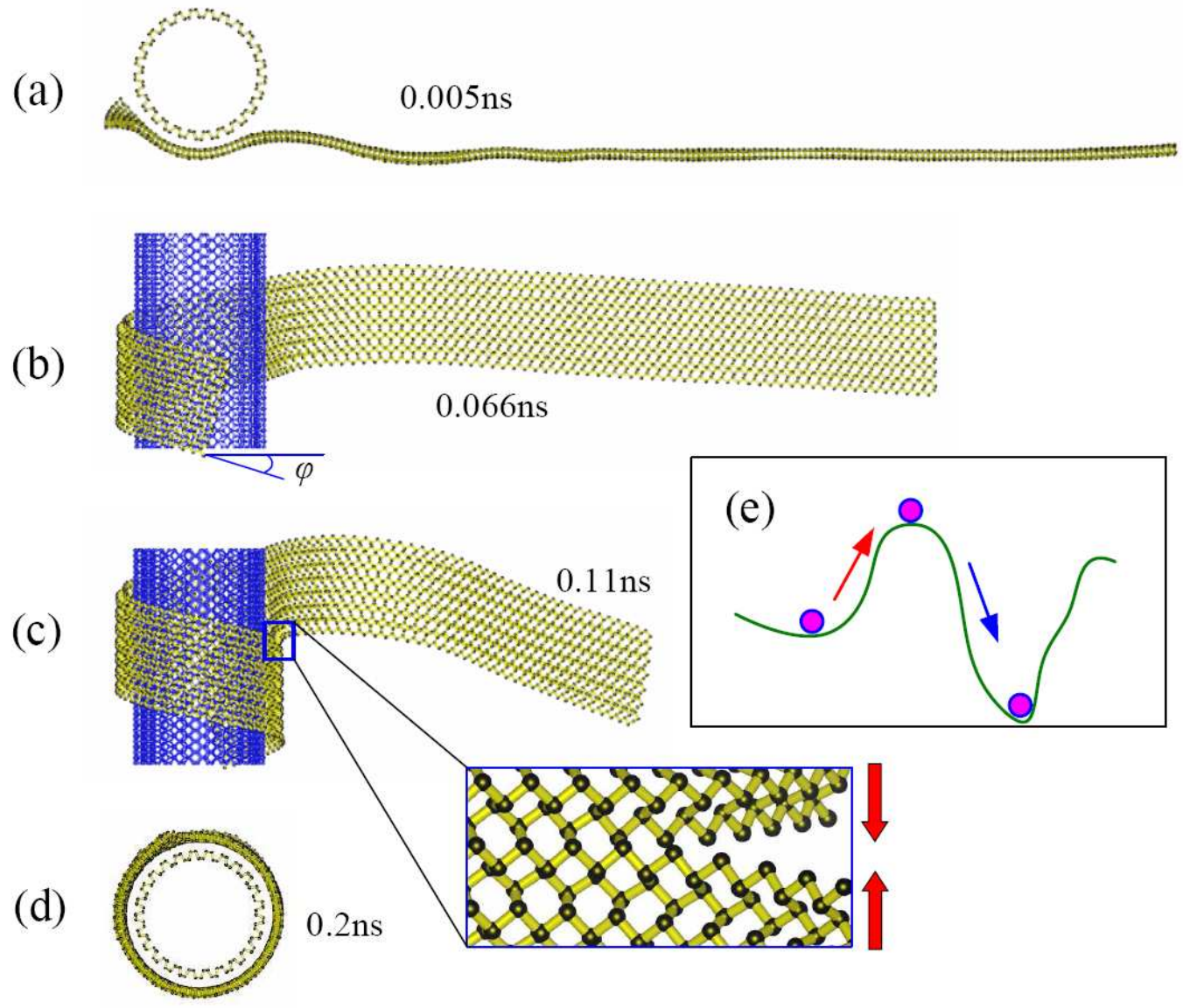

**Fig. 2.** (Color online) Thermodynamic process. Snapshots of self-assembly process of zPNR ($\delta$=13.86) on aPNT ($n$=25) at four different moments (see also Movie1.avi). Chiral angle $\varphi$ in (b) is formed gradually at this stage. The enlarged part in (c) displays bonding process in zPNR edges touched each other. (e) Schematic diagram for van der Waals (vdW) force induced local thermal energy overcoming energy barriers from SW potential.

Prior to simulations, the initial aPNT and zPNR equilibrium configurations were both achieved via the conjugate-gradient (CG) algorithm. Then, the aPNT-zPNR system under free boundary conditions was relaxed to reach a stable thermodynamic state in a canonical (NVT) ensemble with a time step of 1 fs for ~0.5 ns in Nose–Hoover thermostat. As done in other works[2,6,9,10,14], for facilitating the discussion, particularly, for focusing on the physical and mechanical nature of the "chiral" self-assembly itself, we presuppose that experiments are performed in vacuum or inert atmosphere to avoid terminating the edges of zPNR due to ambient oxygen or humidity in practice[27].

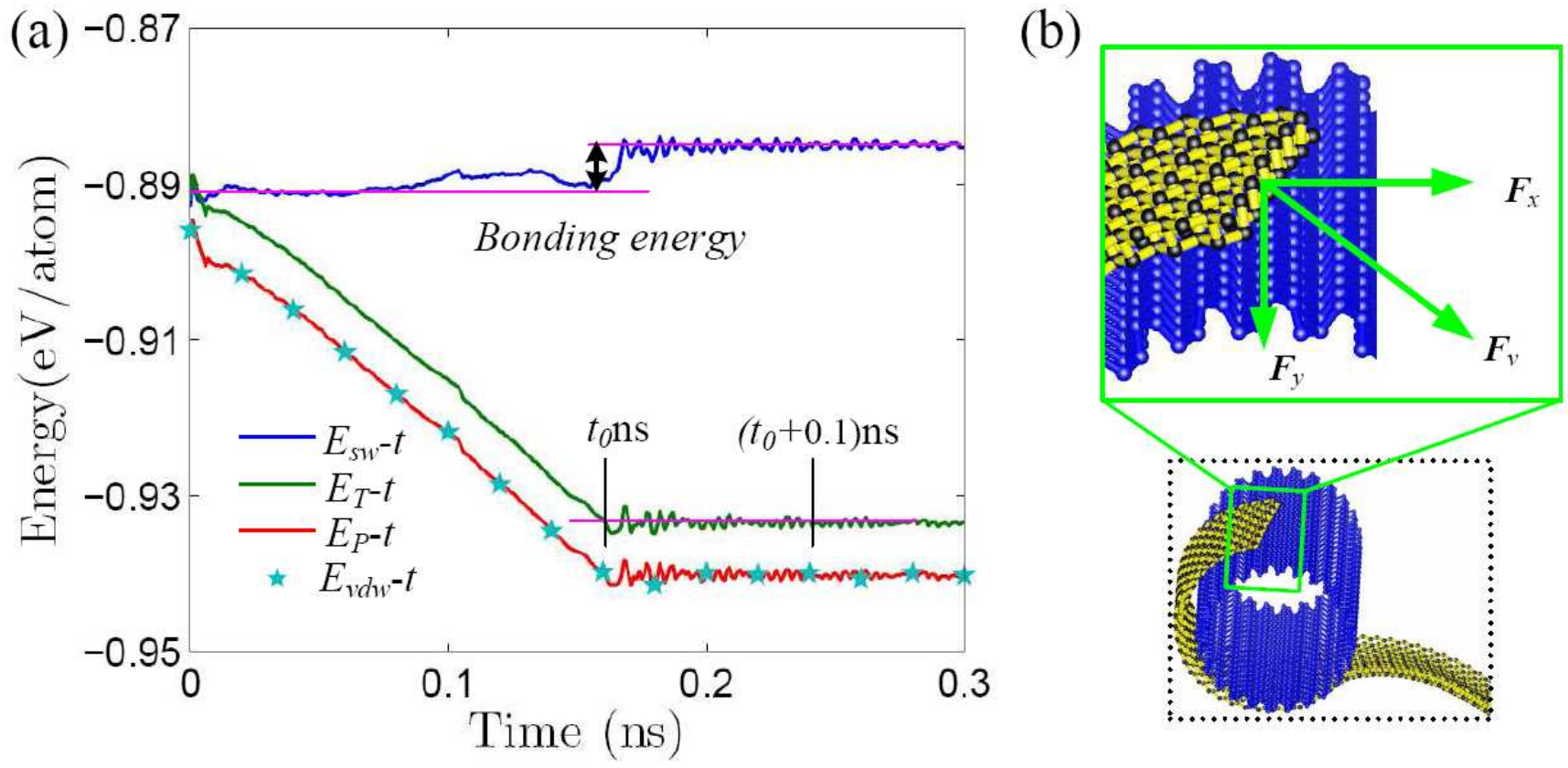

**Fig. 3.** (Color online) Energy feature and free-body diagram. (a) Variation of the (sw/total/potential/vdW) energy with respect to simulating time for the thermodynamic process of aPNT($n$=25) interacting with zPNR($\delta$=13.86). Stable total energy level is reached at $t_0$ ns and a stable chiral tube can be extracted from the simulating system at ($t_0$+0.1)ns. (b) Free-body diagram for nanoribbon rolled on the outer wall of template tube.

Figure 2 shows snapshots at four successive stages of the self-assembly process of a zPNR with aspect ratio $\delta$=13.86 on a fixed aPNT at a constant environmental temperature 30K. Initially, the whole zPNR is at an activated state accompanying slightly bending, twisting, shrinking and expanding due to the thermal motion at a finite environmental temperature as shown in Fig. 2a. As the ribbon sticking the fixed tube, its left side starts to slide and rolls on the outside of the nanotube due to the vdW interaction from the aPNT with the remaining part keeping swinging as shown in Fig. 2b. Interestingly, the nano-ribbon can rearrange itself from the initial state with the longitudinal direction perpendicular to the axis of aPNT as shown in Fig. 1 to form a spiral helix angle and starts to roll around the nano-tube, which is much different from the self-assembly of GNR[12,14,15]. At the same time, the partly-rolled phosphorene ribbon slides much quickly and reciprocally along the tube as the rolling process. A new perfect PNT with certain chirality (which has been actually formed at the second stage shown in Fig. 2b under the given condition) is formed gradually due to the coalescence of connectors as more part of the ribbon is rolled on the aPNT as shown in Figs. 2c-d. The whole process has been videotaped as Movie1.avi in our Supporting Materials, which illustrates the dynamics better than structural snap shots.

The above thermodynamic self-assembly is driven by the minimization of free energy as shown in Fig. 3a, where the total, potential and vdW energy are depicted against the simulation time. All these energies decrease gradually with respect to time and reach a stable minimum (at $t_0$ ~ 0.16 ns as denoted in the figure for the total energy), which is similar to that of GNR[2,12,14,15]. The free energy of zPNR self-assembly after equilibrium is much larger than that of GNR[14,15,16], since the configuration energy of graphene is -9.214eV/atom, which is much smaller than that of phosphorene, -5.365eV/atom, according to density functional theory (DFT) calculations[3,5,22]. The potential curve is coincided with that of the vdW interaction indicating that the self-assembly process is actually driven by the vdW force. Figure 3b presents a free-body diagram for the part of raw ribbon rolled on the surface of template tube and the traction force $\boldsymbol{F}_v$ acting on the nano-ribbon from vdW interaction, which can be decomposed into the component $\boldsymbol{F}_x$ at a tangent of nano-tube pulling the ribbon to roll around the ring, and the component $\boldsymbol{F}_y$ along the axis of nano-tube pulling the ribbon to slide along the groove of corrugated crystal.

**Table 1.** Comparisons between theoretical formulas and MD simulation for results of the chiralities of self-assembled cPNT by zPNR on aPNT.

| zPNR($L_y$/nm) | cPNT [$(n,m_c)$ \| (φ/°)] | | aPNT($r$/nm) |
|---|---|---|---|
| [$h$=0.21nm] | Formula (3)\| (4) | MD-simulation | [$\sigma$=0.35nm] |
| 1.84($\delta$=24.25) | (4,31) \| 10.24 | (4,30) \| 10.58 | 1.10($N_L$ =30) |
| 1.84($\delta$=48.25) | (4,45) \| 7.10 | (4,45) \| 7.10 | 1.84($N_L$ =30) |
| 3.23($\delta$=20.71) | (7,37) \| 14.84 | (7,37) \| 14.84 | 1.47($N_L$ =30) |
| 3.23($\delta$=20.71) | (7,44) \| 12.56 | (7,44) \| 12.56 | 1.84($N_L$ =20) |
| 3.23($\delta$=13.86) | (7,44) \| 12.56 | (7,44) \| 12.56 | 1.84($N_L$ =30) |

Note that: $\sigma$=0.35 nm is an average relaxed distance between phosphorene ribbon and phosphorus nanotube, slightly different from that of bilayer phosphorene.

The chiral vector of the chiral phosphorus nanotube generated can be described theoretically as

$$\boldsymbol{C}_h = n\boldsymbol{a}_1 + m_c\boldsymbol{a}_2 = n\boldsymbol{a}_1 + \left[\frac{\sqrt{(\pi d_c)^2 - (n|\boldsymbol{a}_1|)^2}}{|\boldsymbol{a}_2|}\right]\boldsymbol{a}_2 \tag{3}$$

where $d_c$=2($r$+$h$+$\sigma$) is the diameter of the cPNT($n$, $m_c$) and [$x$] denotes the nearest integer of $x$. Note that more precise lattice constants based on the DFT results $|\boldsymbol{a}_1|$=0.462 nm and $|\boldsymbol{a}_2|$=0.330 nm are adopted for convenience of theoretical predictions[5], which are in agreement with that obtained from our MD calculations (the average lattice constants are 0.441nm and 0.331nm, respectively). According to the Eq. (3), the chiral angle in Fig.2b can be written as

$$\varphi = \frac{180}{\pi}\arctan\left[\left(n\left|\boldsymbol{a}_1\right|\right)/\left(m_c\left|\boldsymbol{a}_2\right|\right)\right] \tag{4}$$

We conducted a series of simulations using the zPNR with varied aspect ratios $\delta$ and the aPNT with different radii $r$ or chiral indexes $n$ to compare the theoretical predictions and MD simulations in which stable chiral tubes can be extracted from the simulating system at ($t_0$+0.1) ns. It is shown from Table 1 that the predictions of chiralities for the produced phosphorus nanotubes by Eqs. (3) and (4) are well consistent with the results of MD simulations, meaning that the chirality of the self-assembled cPNT is dependent on the width $L_y = n|\boldsymbol{a}_1|$ of zPNR and the radius $r$ of aPNT.

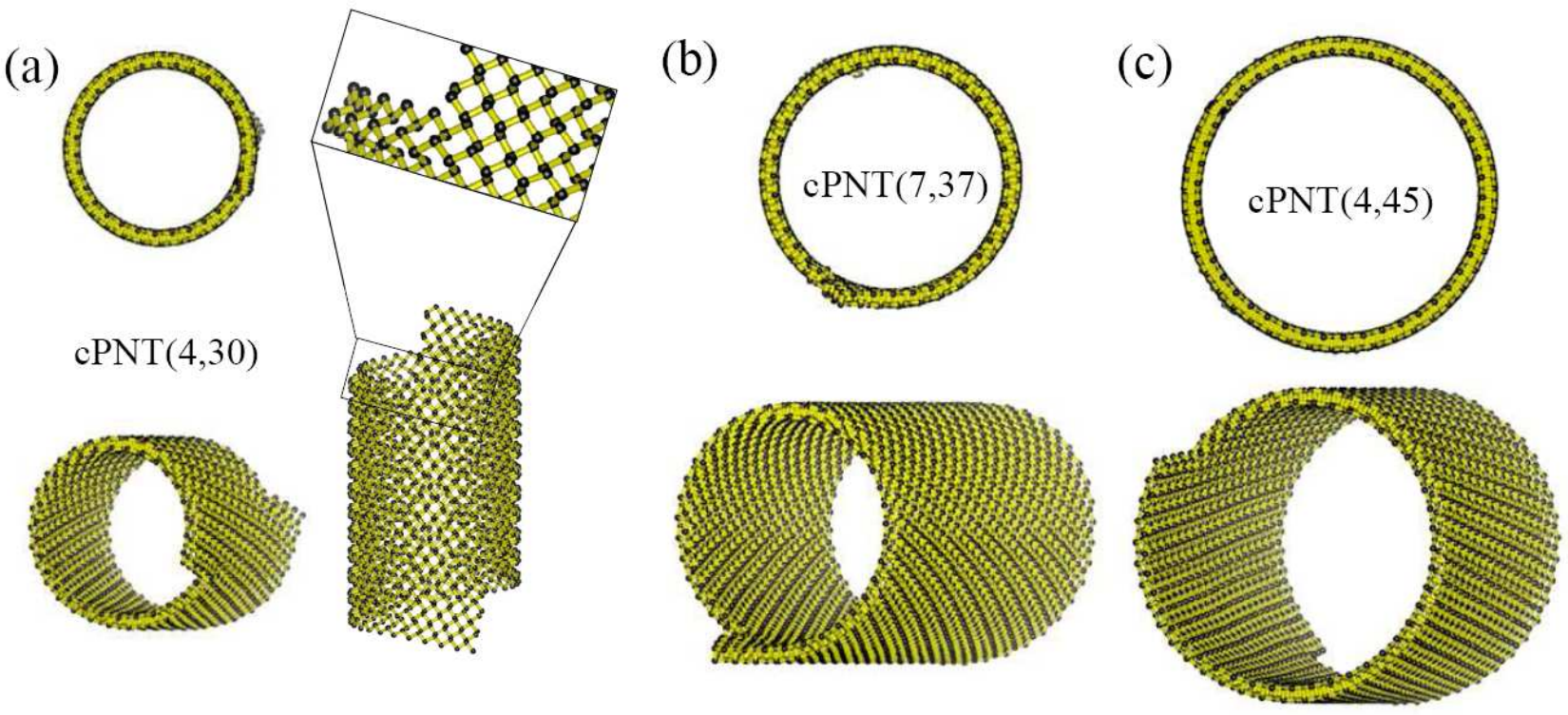

**Fig. 4.** (Color online) Three typical cPNTs with isometric and top views: (a) cPNT(4,30), (b) cPNT(7,37), (c) cPNT(4,45), whose self-assemblies are induced by the aPNTs with $n$=15, 20 and 25, respectively. The enlarged part in (a) displays the bonding structure in cPNT(4,30).

Figure 4 presents three typical cPNTs with front and top views, whose radii $r_c$ are 1.66, 2.02, 2.39nm, respectively. The enlarged part in Fig.4a as an example displays the perfect bonding structure in cPNT(4,30). Such a mutual bonding was also discussed by Czerwinski, *et al.* based on classic AIREBO force field[28] and so was it for Tersoff potential in some specific conditions (such as high pressure)[21, 22,29]. But in current case, it is not only attributed to the dangling chemical bonds in zPNR edges, but also due to the strong constraining force from vdW interaction overcoming the activation energy barriers from SW potential[21,23]. Figure 3e presents a schematic diagram, where vdW force induced local thermal energy make the bonding edges climb to a high energy barrier of SW potential, and then drop to a lower energy well. Therefore, as just shown in the detailed illustration of Fig.2c, the helical pitch becomes so small that the zPNR edges can squeeze and bond with each other. It should be noted that, the strong impact can be essentially attributed to an extremely high pressure, and the bonding just forms by means of the mathematical form of non-mechanical (SW) potential functions consisting of two- and three-body interactions[21,22,23,29]. It is also noted that some mechanical classical force fields such as CHARMM, can not arrived at bond breaking or forming even in current case, where bonds, angles and dihedrals are all assigned according to the (phosphorene) crystal configuration[14,21,22]. Morevoer, Figure 3a presents a variation of SW energy against the simulation time. It can be seen that the energy curve almost hold a level value before the third stage and after the forth stage of the winding process except tiny fluctuations due to small local distortions of the zPNR. A general increasing energy is observed from ~0.09 to ~0.16ns with evident fluctuations. It shows that an average bonding energy is saved in the new tube as an intrinsic SW potential during the bonding process, accompanying by large local distortions. The single bond forming can be roughly estimated as a local thermal energy at extremely high temperature of 29158K ($k_B T$ = 0.005 eV/atom in Fig.3a with total 5025 atoms in the system and assuming that 10 atoms forming bond at the same time). This further demonstrates a fully admissible bonding condition. Comparing the cPNT(4,30) in Fig.4a with the cPNT(4,45) in Fig.4c for the same chirality index $n$=4, it is known that the chirality of son-tube is independent on the length $L_a$ (also, $N_L=L_a/|\boldsymbol{a}_2|$ as an integer indicator listed in Table 1) of parent-tube and/or the length $L_x$ of raw-ribbon, but its length can be measured by them fully, and could approximately conform to the theoretical formula

$$L_c \approx \left(\frac{m}{m_c}-1\right)n\left|\boldsymbol{a}_2\right|\sin\left(\frac{\varphi\pi}{180}\right) \tag{5}$$

For the simulated cPNT(7,44) in Table 1 with the same width $L_y$=3.23nm of zPNRs and the same radius $r$=1.84nm of the aPNT, according to Eq. (5) the effective lengths $L_c$ are calculated as 0.60nm and 1.15nm, respectively, also in good agreement with the results of MD simulations.

Due to the anisotropic nature of phosphorene and the bending instability along its zigzag direction[5,10], the zPNRs prefer moving forward in a spiral way outside the aPNTs to bend themselves directly, and finally form chiral tubes with different radii, as shown in Fig.4a-c. Recently, Fernández-Escamilla *et al.* pointed out based on their DFT calculations that cPNTs hold larger binding energy than zPNTs and the values are comparable to aPNTs[20]. For checking the stability of the assembled cPNT, the Supporting Material Movie2.avi presented the thermodynamic process of the cPNT(7,37) shown in Fig.4b at 100K and the asymmetric ends was directly chopped off (e.g., a recent controlled sculpture method of phosphorene nanoribbons developed by Das *et. al.*[30]) to eliminate the sway induced the stress concentration. The result shows that the chiral tube can hold at least 0.4ns at 100K in a canonical ensemble, much more stable than the zPNT[10]. Also, the large thermal motion of the cPNT(7,37) is constrained by the SW potential to be a corresponding thermal deformation and partially heat-driven oscillations.

In order to further study the temperature dependence and size effect, it is necessary to introduce a self-assembly energy $E_{es}$ and velocity $V_{es}$ as

$$E_{es} = E_s - E_e \quad and \quad V_{es} = \frac{L_x}{t_e - t_s} \tag{6}$$

where $E_s$ and $E_e$ are the initial and final total energy, and $t_s$ and $t_e$ are the start and finish time of the self-assembly with $t_s$=0 and $t_s=t_0$+0.1, as noted in Fig.1a. The investigated temperature is kept at $T$<100K or one third of room temperature during simulations, which can be viewed as an intermediate temperature condition to avoid instability of the formed chiral tube as stated above[6,9,10]; meanwhile, too low temperature is not allowed owing to no enough kinetic energy moving the raw ribbon and pushing it sticking the template tube.

Figure 5a shows that the self-assembly energy and velocity change with aspect ratio for the self-assemblies induced by the aPNT($n$=25) and the self-assembly energy increases while the self-assembly velocity decreases with the aspect ratio, regardless of the temperature being at 10 or 30K. Moreover, for the aPNT($n$=25) induced self-assembly of the zPNR($\delta$=13.86), as presented in Fig.5b, the self-assembly energy decreases while the self-assembly velocity increases with the temperature. This is because the higher temperature mainly contributes the kinetic energy and makes the average atom velocity quicker to speed up the self-assembly process. Under given temperature and length of one side, the smaller aspect ratio mainly lowers the potential energy difference shortening the self-assembly time and makes the rolling easier on the outside surface of the aPNT[14,21]. At the temperature of 70K or higher, the self-assembly process becomes so fierce that a fracture behavior happens, as shown in the inset of Fig.5b for details, and this further demonstrates the controlled temperature should be kept at $T$<100K.

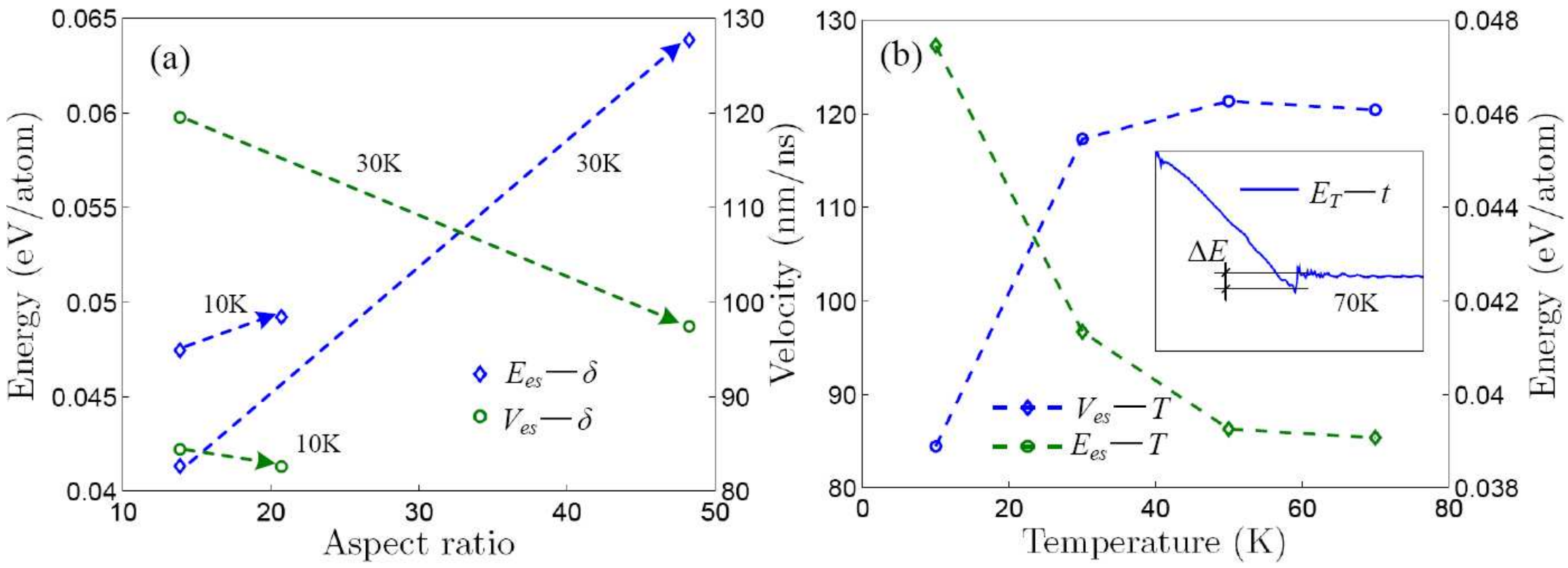

**Fig. 5.** (Color online) Self-assembly energy and velocity curves. Self-assembly energy and velocity change with (a) aspect ratio for self-assembly induced by aPNT($n$=25) and (b) Kelvin temperature for self-assembly of zPNT($\delta$=13.86) induced by aPNT($n$=25). The inset in (b) shows variation of total energy with respect to time at 70K, in which a fracture energy $\Delta E$ is defined approximately, being the same as the average bonding forming energy presented in Fig.3a.

In summary, a self-assembly mechanical behavior from phosphorene to chiral phosphorous nanotube was studied systematically for the first time, based on classic molecular dynamics simulations incorporating theoretical analysis. We predicted that a template armchair phosphorus nanotube can induce a zigzag phosphorene nanoribbon to form a corresponding chiral phosphorous nanotube. The van der Waals interaction between the template nanotube and raw nanoribbon is the driving force of self-assembly process, which leads to the bending, twisting and shrinking of the raw ribbon in vacuum and sticking, sliding and rolling outside the template tube. Particularly, the interaction is transformed to the intrinsic deformed and chemical bonding energies of the synthesized tube after overcoming the energy barriers from Stillinger-Weber potential. A positive correlation was found between the self-assembly energy and aspect ratio or between the self-assembly velocity and temperature, while a negative correlation was found between the self-assembly energy and temperature or between the self-assembly velocity and aspect ratio. The formed chiral tube is thermodynamically stable and its chiralities and sizes can be controlled pretty well.

The work reported here is supported by the NSFC of China (Grant Nos. 11602270, 11574029, and 11532013), the MOST of China (Grant No. 2014CB920903, 2016YFA0300600), and the National Basic Research Program of China (“973” Project, Grant No.2012CB937500). The first author thanks Dr Dongmei Hu in Jilin University for helpful discussions and the reviewer from ACS Nano for his/her constructive suggestions on classic bonding mechanism.